\documentclass{PoS}
\usepackage{amsmath}
\usepackage{wrapfig}

\title{String tension from smearing and Wilson flow methods}

\ShortTitle{String tension from smearing and Wilson flow methods}

\author{Antonio Gonz\'alez-Arroyo$^{ab}$ \\
\llap{$^a$}Instituto de F\'{\i}sica Te\'orica UAM/CSIC, C/ Nicol\'as Cabrera 13-15\\
       Universidad Aut\'onoma de Madrid, E-28048--Madrid, Spain \\
\llap{$^b$}Departamento de F\'{\i}sica Te\'orica, C-15 \\
       Universidad Aut\'onoma de Madrid, E-28049--Madrid, Spain\\
E-mail: \email{antonio.gonzalez-arroyo@uam.es}}       
\author{\speaker{Masanori Okawa}$^c$ \\      
\llap{$^c$}Graduate School of Science, Hiroshima University\\
Higashi-Hiroshima, Hiroshima 739-8526, Japan\\
E-mail: \email{okawa@sci.hiroshima-u.ac.jp}}

\abstract{Recently, we proposed a new method to extract the string tension from
4-dimensionally smeared Wilson loops.  In this talk,
we first show that the results obtained using this smearing method are
identical  to those obtained by Wilson flow, once the time step is 
sufficiently small.
We then demonstrate the practical advantage of our method by applying it to
the calculation of string tension in SU(3) Yang-Mills theory. }

\FullConference{The 32nd International Symposium on Lattice Field Theory,\\
		23-28 June, 2014\\
		Columbia University New York, NY}

\begin{document}

\section{Introduction}

\vspace{-0.2cm}

It is well known that Wilson loops of large size are quite noisy.  
The usual way to circumvent this difficulty is to apply 3-dimensional smearing 
to spatial link variables, and string tensions are then extracted from 
the smeared $q {\bar q}$ potential \cite{3dsm}.
Recently, we proposed a new method to calculate string tension from 4-dimensionally
smeared Wilson loops \cite{4dsm1,4dsm2}.  The string tension is extracted directly from
the continuum limit of Creutz ratios.

Recently, Wilson flow became a standard technique to regularize local operators 
in which local link variables are evolved along fictitious time $t$ \cite{wilsonf}.   Link variables 
at $t$ are essentially the average of the original link variables in a spherical 
region of size $\sqrt{8t}$ in 4-dimensional space-time.  Thus, by definition, 
Wilson flow is closely related to the 4-dimensional smearing method \cite{NN}.       

In this talk we first explain our 4-d smearing method, and then demonstrate that
the same  results are  obtained when using  Wilson flow instead,  provided that the time step $\Delta t$ of 
the numerical time evolution is sufficiently small.   Previously the
method was used in the context of extracting the string tension for
the large N  limit of SU(N) gauge theory, and comparing it with the
result of reduced models~\cite{TEK1,TEK2}.  Here we will apply it to  SU(3) Yang-Mills theory.
To test the scaling of the results we will study  three  different
values of the coupling while  keeping constant the physical size of the box. 
Apart from the string tension we will also obtain a continuum function
built in terms of Wilson loops. We will then  compare our results
to those obtained from spatially smeared $q {\bar q}$ potential.

\vspace{-0.4cm}

\section{4-dimensional Ape smearing}

\vspace{-0.3cm}

We consider 4-dimensional Ape smearing defined by \cite{APE,NN}

\vspace{-0.3cm}

\begin{equation}
 U_{n,\mu}^{smeared}={\rm Proj}_{SU(3)} \left[ (1-f) U_{n,\mu} 
+ {f \over 6}  \sum_{\nu \ne \mu=\pm1}^{\pm4}U_{n,\nu} U_{n+\nu,\mu} U_{n+\mu,\nu}^\dagger \right], 
\end{equation}    

\noindent
where ${\rm Proj}_{SU(3)}$ stands for the operator for projection onto the SU(3) matrices. 
As explained in Ref.~\cite{NN}, for sufficiently small steps  $\Delta t
= f/6$ the smearing procedure should scale with the variable $t=f n_s /6$,
where  $n_s$ denotes the number of smearing steps. In the next
section, we will show that the same results are obtained from Wilson
flow at the fictitious time equal to $t$.

\begin{figure}[t]
 \begin{minipage}{0.5\hsize}
  \begin{center}
   \includegraphics[width=65mm]{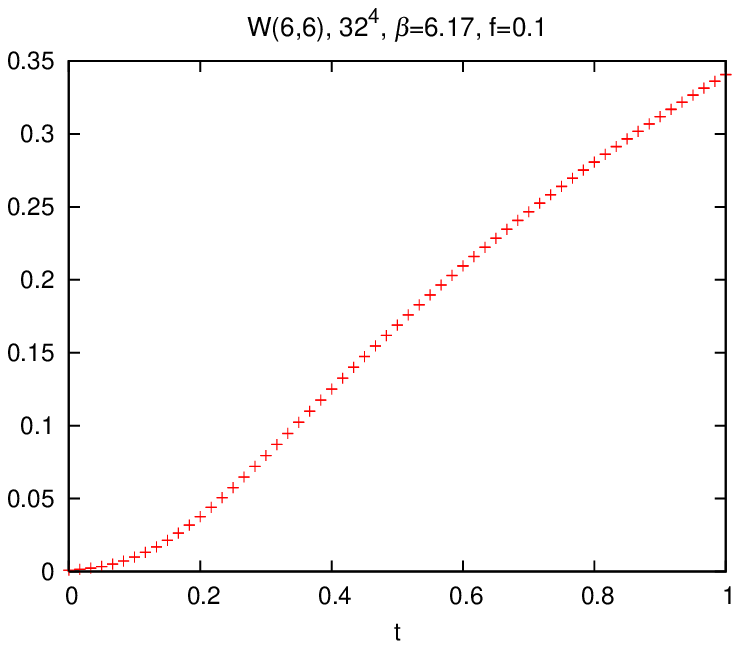}
  \end{center}
\vspace{-0.7cm}  
  \caption{$t$ dependence of Wilson loop $W(6,6)$.}
  \label{fig_wilson_6x6}
 \end{minipage}
 \begin{minipage}{0.5\hsize}
  \begin{center}
   \includegraphics[width=65mm]{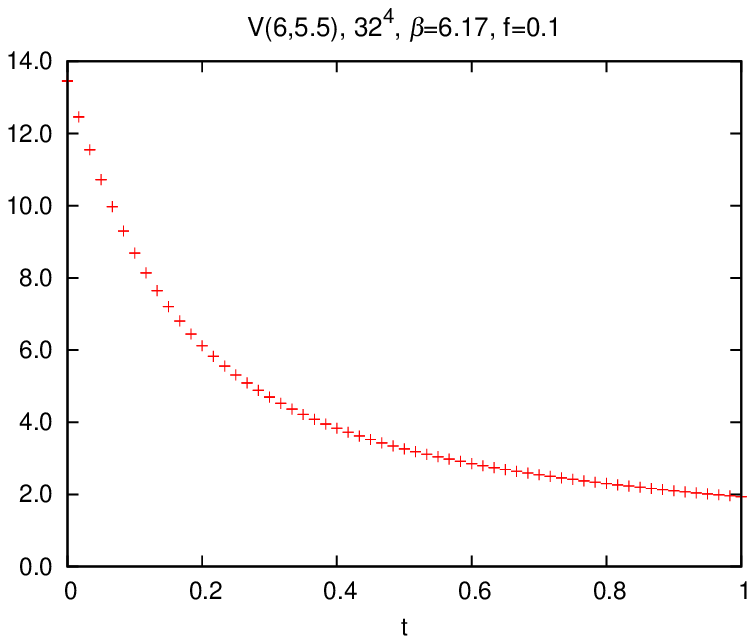}
  \end{center}
\vspace{-0.7cm}  
  \caption{$t$ dependence of $q {\bar q}$ potential $V(6,5.5)$.}
  \label{fig_potential_6x5.5}
 \end{minipage}
\end{figure}

In the literature, it is customary  to apply 3-dimensional Ape smearing to spatial link variables
to obtain smeared  $q {\bar q}$ potential with small noises \cite{3dsm}.
On the other hand, 4-dimensional Ape smearing is not extensively used so far \cite{LN},  because 
smeared Wilson loops and smeared $q {\bar q}$ potentials have huge $t$ dependences.
As an example, In Fig. \ref{fig_wilson_6x6}, we show the $t$ dependence of a Wilson loop and 
in Fig. \ref{fig_potential_6x5.5}, that of the discretized $q {\bar q}$ potential  $V(R,T)= -\log\left[ W(R,T+1/2) / W(R,T-1/2) \right]$  
with integer $R$ and half-integer $T$.   The simulation has been made on a $32^4$ lattice 
at $\beta=6.17$ with $f$=0.1.
The smeared Wilson loops can be considered regulated versions with $t$
playing the role of an ultraviolet cut-off. The divergences of these 
quantities imply then a strong $t$ dependence as we approach the
continuum limit. 

Our proposal is rather  to consider Creutz ratios

\begin{figure}[b]
 \begin{minipage}{0.5\hsize}
  \begin{center}
   \includegraphics[width=65mm]{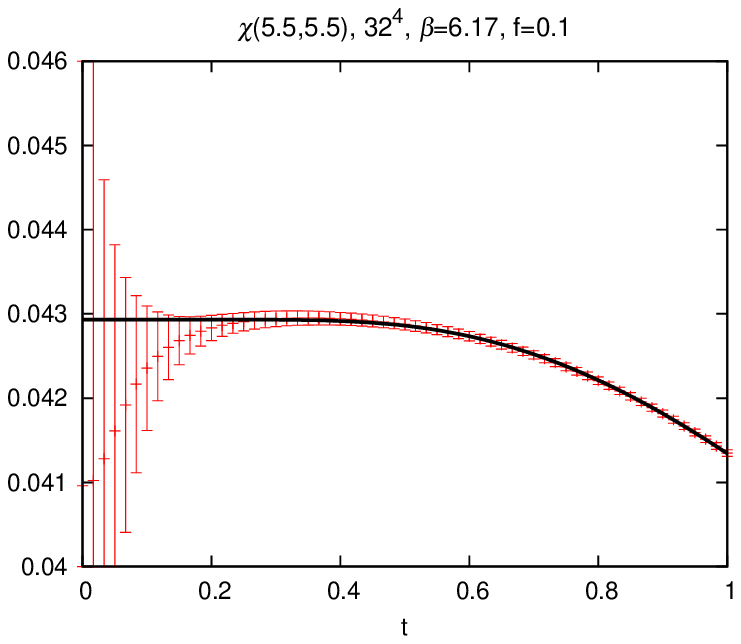}
  \end{center}
\vspace{-0.7cm}
  \caption{$t$ dependence of the 4-d smeared Creutz \hspace{2mm} \newline
ratio $\chi(5.5,5.5)$.  
The solid line is the fit with (2.3).}
\label{fig_cratio_5.5x5.5} 
 \end{minipage}
 \begin{minipage}{0.5\hsize}
  \begin{center}
   \includegraphics[width=65mm]{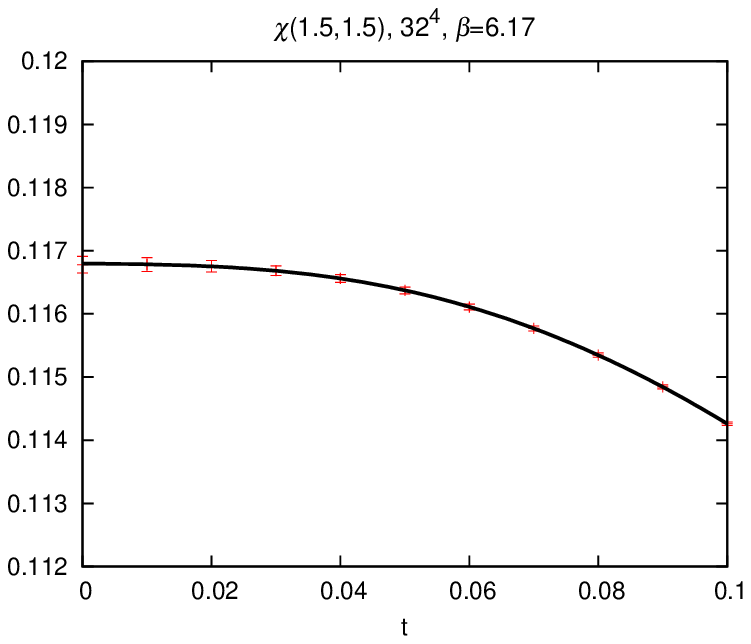}
  \end{center}
\vspace{-0.7cm}  
\caption{$t$ dependence of the 4-d smeared Creutz \hspace{2mm} \newline
ratio $\chi(1.5,1.5)$.  
The solid line is the fit with (2.3).} 
\label{fig_cratio_1.5x1.5} 
 \end{minipage}
\end{figure}

\vspace{-0.2cm}

\begin{equation}
\chi(R,T)= -\log{W(R+1/2,T+1/2) W(R-1/2,T-1/2) \over W(R+1/2,T-1/2) W(R-1/2,T+1/2)}
\end{equation}

\noindent
with half-integer $R$ and $T$. Since Creutz ratios $\chi(R,T)$ are free from ultraviolet divergences,
we expect that the smeared  quantities have a well defined
$t\rightarrow 0$ limit.  As an example,  in fig. \ref{fig_cratio_5.5x5.5} we show $\chi(5.5,5.5)$    
as a function of $t$ at $\beta=6.17$ with $f$=0.1. Notice however that,
for small values of $t$, $\chi(R,T)$ has huge errors. This is
understandable since the Creutz ratios are constructed in terms of the
ultraviolet divergent quantities which are affected by large errors.
On the other hand as we increase $t$, the error drops considerably. 
The $t$-dependence of Creutz ratios is understandable and computable
in perturbation theory. Restricting ourselves to not too large values
of $t$ this suggests a parameterization of the type

\vspace{-0.5cm}

\begin{equation}
\chi(t)= a \left\{ 1 - \exp\left({ -b \over t+c} \right) \right\}.
\label{eq_fit_function}
\end{equation}

\noindent 
which, as seen in fig.~\ref{fig_cratio_5.5x5.5}, perfectly fits the
data. It is clear that around $t\sim 0.4$ the curve is essentially horizontal 
and its value gives a good estimate of the $t\rightarrow 0$ limit. 
However, to avoid fitting a constant within a subjective range, it is
much better to use the previous functional form and fit also the 
values for $t>0.5$.

For small loop sizes, Creutz ratios generally have small statistical errors. 
As an example, we show in fig. \ref{fig_cratio_1.5x1.5} the $t$ dependence of 
$\chi(1.5,1.5)$.  The necessity of using smearing is then not clear. 
However, notice that the fit with (\ref{eq_fit_function}) is indeed
quite good.    It should be noted that, for small loop sizes, the existence 
of the parameter $c$  is crucial to fit the data. The smeared Creutz ratio 
monotonically decreases as  we increase $t$, implying that there is actually 
no plateau.

\vspace{-0.8cm}

\section{Relation with Wilson flow}  

\vspace{-0.4cm}

\begin{figure}[htbp]
\begin{minipage}{0.5\hsize}
\begin{center}
   \includegraphics[width=65mm]{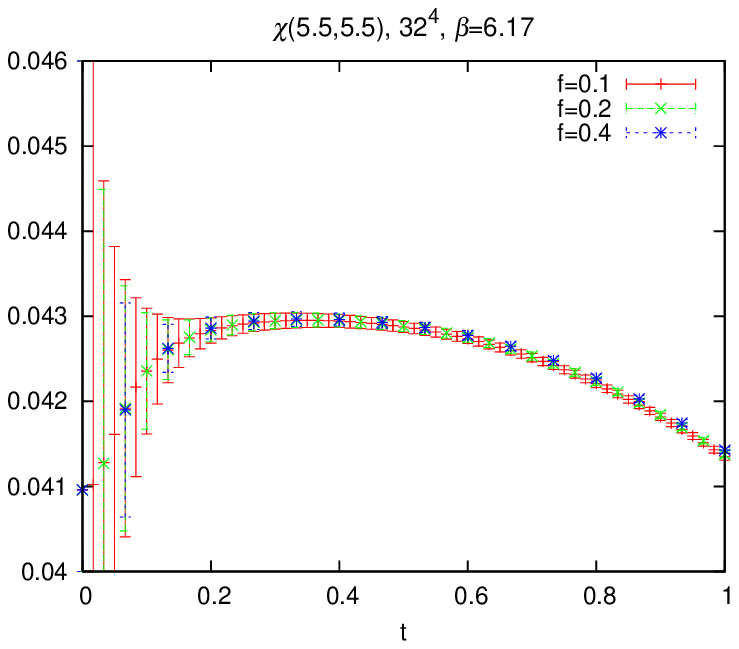}
\end{center}
\vspace{-0.7cm}  
  \caption{Smeared Creutz ratio with $f$=0.1, 0.2 \hspace{1mm} \newline and 0.4.}
  \label{fig_cratio_f}
 \end{minipage}
 \begin{minipage}{0.5\hsize}
  \begin{center}
   \includegraphics[width=65mm]{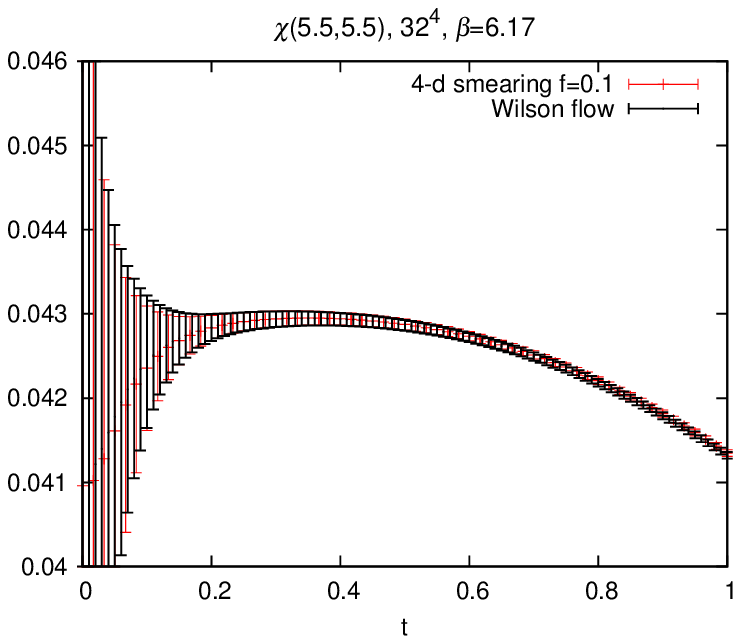}
  \end{center}
\vspace{-0.7cm}  
\caption{$\chi(5.5,5.5)$ from 4-d smearing (red symbol) and that from Wilson flow (black symbol).}
\label{fig_smearing_wilson}
 \end{minipage}
\end{figure}

As mentioned earlier the smearing dependence is actually better
expressed in terms of $t$. This is clearly seen in 
Fig.~\ref{fig_cratio_f} where results for three different values of 
$f$ are compared. We also analyzed the dependence of the result on the 
regulating method. For that purpose we applied  Wilson flow with 
fictitious time $t$, defined by \cite{wilsonf}

\vspace{-0.4cm}

\begin{equation}
dV_{n,\mu}(t) / dt  = -g_0^2 \left\{ \partial _{n,\mu} S_W \right\} V_{n,\mu}(t), \ \ \ V_{n,\mu}(t=0)=U_{n,\mu}.
\end{equation}

\begin{wrapfigure}{t}{55mm}
\begin{center}
\includegraphics[width=55mm]{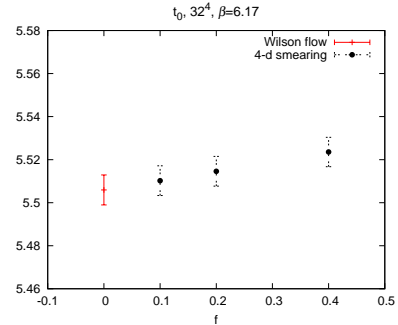}
\end{center}
\vspace{-0.7cm}
  \caption{$f$ dependence of $t_0$.}
  \label{fig_t_0}
\end{wrapfigure}

\noindent 
We have used third-order Runge-Kutta scheme with $\Delta t = 0.01$.  
In fig. \ref{fig_smearing_wilson}, we compare the results 
obtained from 4-d smearing ($f$=0.1) with that from Wilson flow.
It is obvious that both methods give identical values of the smeared
Creutz ratios once the time step  $f$ is sufficiently small.  
A slight $f$-dependence is observed when computing the scale $t_0$
defined by   $\left\{ t^2 \left< E \right> \right\} _{t=t_0}$ = 0.3
\cite{wilsonf}, with $E$ evaluated from the lattice version of $F_{\mu\nu}$.
The result is displayed in Fig.~\ref{fig_t_0} and compared with the
result obtained by Wilson flow. These slight discretization errors,
however, do not affect the $t \rightarrow 0$ extrapolation of
Creutz ratios.  

Concerning the dependence of the parameters of the fit on the size of
the loop, perturbation theory predicts that $b \sim 25/R^2$. Indeed, 
our data  agrees with  a dependence on $25/(R-0.5)^2$, which is
consistent with this prediction up to discretization effects.

\vspace{-0.5cm}

\section{Application to SU(3) Yang-Mills theory}  

\vspace{-0.4cm}

In order to demonstrate the practical usefulness of our method, we have applied it 
to the SU(3) Yang-Mills theory and calculated the string tension and
the continuum version of the Creutz ratios for that case.
To monitor scaling behaviour  we have made simulations at three lattices having 
approximately equal physical  size as shown in Table 1.  The scale
${\bar r}$ was defined in Refs.~\cite{4dsm1,4dsm2} and its definition will be
given later.
\noindent
In this talk, we will concentrate on the diagonal Creutz ratios $\chi(R,T)$ with $R=T$,
although there are a lot of interesting physics in off-diagonal $\chi(R,T)$ \cite{4dsm1,4dsm2}.

\begin{table}[htb]
\begin{center}
\begin{tabular}{|c c c c c c|} \hline
  Lattice  & $\beta$ & $N_{cnfg}$ & $a/{\bar r}$ & $t_0/a^2$ & $\sqrt{8t_0}/{\bar r}$ \\ \hline 
    $24^4$ &  5.96   & 1000       & 0.2117(5)    & 2.794(3) & 1.001(2) \\ \cline{1-6}
    $32^4$ &  6.17   &  300       & 0.1499(2)    & 5.506(7) & 0.995(2)\\ \cline{1-6}    
    $48^4$ &  6.42   &  100       & 0.1052(3)    & 11.17(3) & 0.994(3)\\  \hline         
  \end{tabular}
\end{center} 
\vspace{-0.3cm}
 \caption{Properties and statistics of three lattice runs. }
\label{table1}
\end{table}

\begin{figure}[htbp]
 \begin{minipage}{0.5\hsize}
  \begin{center}
   \includegraphics[width=70mm]{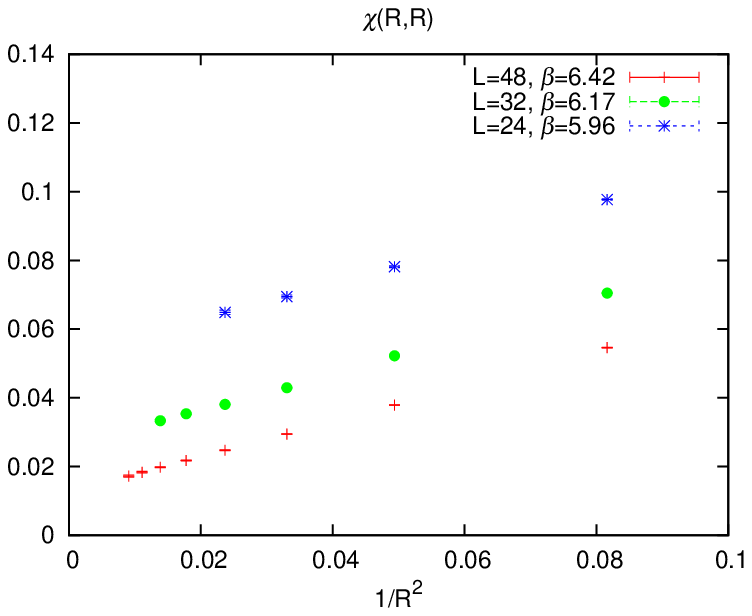}
  \end{center}
\vspace{-0.5cm}  
  \caption{Creutz ratio $\chi(R,R)$ as functions of $1/R^2$.}
  \label{fig_cratio_lattice}
 \end{minipage}
 \begin{minipage}{0.5\hsize}
  \begin{center}
   \includegraphics[width=70mm]{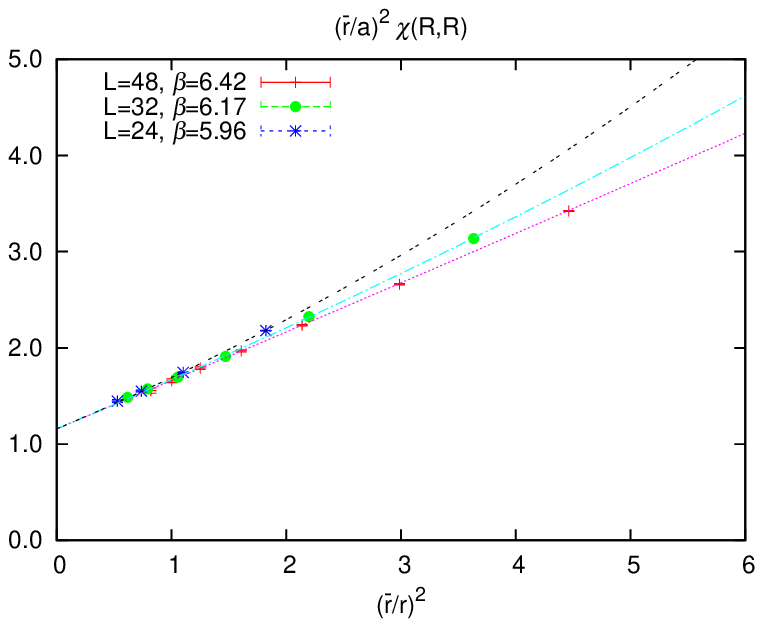}
  \end{center}
\vspace{-0.5cm}  
  \caption{ $({\bar r}/a)^2 \chi(R,R)$ as functions of $({\bar r}/r)^2$.}
  \label{fig_cratio_phys}
 \end{minipage}
\end{figure}

In Fig. \ref{fig_cratio_lattice}  we display the results obtained for
the diagonal Creutz ratios $\chi(R,R)$ using the 4d smearing method
described earlier. The statistical error are too small to be seen at 
the  scale of the plot. To analyze scaling it is better to
consider the quantity $R^2 \chi(R,R)$ which in the continuum limit
tends to a dimensionless function as follows:

\vspace{-0.3cm}

\begin{equation}
R^2 \chi(R,R) \stackrel{a \rightarrow 0}{\longrightarrow} r^2
{\tilde F}(r) + \mathcal{O}(\frac{a^2}{r^2})
\end{equation}

\vspace{-0.1cm}

\noindent
where $r=R a$. The function ${\tilde F}(r)$ can be written in terms of
Wilson loops as follows:

\vspace{-0.2cm}

\begin{equation}
{\tilde F}(r) = -\lim_{t \rightarrow 0} \left. \frac{\partial^2 \log W_t(r,r')}{\partial r \partial
r'}\right|_{r=r'} .  
\end{equation}

\vspace{-0.1cm}

\noindent
where  $W_t(r,r')$ is the continuum Wilson loop regulated with
smearing parameter $t$. Although  $\lim_{t \rightarrow 0}
W_t(r,r')$ is not well-defined, it is in the expression of ${\tilde
F}(r)$.  The scale $a(\beta)$ can be fixed by any method. However, 
one can use the information of the Creutz ratios themselves to obtain
them in units of a new scale ${\bar r}$ (\'a la Sommer), defined  by  the equation ${\bar r}^{\ 2}
\tilde{F}(\bar{r})=1.65$~\cite{4dsm1,4dsm2}. 

In applying the previous ideas to our data, we first realize that, in
the range of $R$ values explored, the Creutz ratios can be perfectly 
fitted by a second degree polynomial in $1/R^2$. Combining this 
with the previous information we might write 

\vspace{-0.3cm}

\begin{equation}
\left( { {\bar r} \over a} \right)^2 \chi(R,R) = \sigma {\bar r}^{\ 2} + 2 \gamma 
\left({{\bar r} \over r}\right)^2 + 4 \left({{\bar r} \over r}\right)^4 \left[ c + d \left({ a \over {\bar r} }\right)^2  \right]
\label{eq_cratio_lat}
\end{equation}

\noindent
in which we have allowed for a term (associated to the $d$ parameter)
to account for the leading correction to scaling effects. Thus the
continuum function ${\tilde F}(r)$ is described by

\vspace{-0.2cm}

\begin{equation}
 {\bar r}^{\ 2} {\tilde F}(r) = \sigma {\bar r}^{\ 2} + 2 \gamma 
\left({{\bar r} \over r}\right)^2 + 4 c \left({{\bar r} \over r}\right)^4 
\label{eq_cratio_cont}
\end{equation} 

\noindent
The definition of the scale  ${\bar r}$ implies a relation among the
parameters as follows 

\vspace{-0.4cm}

\begin{equation}
 {\bar r}^{\ 2} {\tilde F}({\bar r}) = \sigma {\bar r}^{\ 2} + 2 \gamma 
 + 4 c = 1.65
\label{eq_new_scale}
\end{equation}

\vspace{-0.2cm}

\noindent
which allow us to determine  $c$ in terms of the other parameters. 
Summarizing, we are led to a simultaneous parameterization of the  Creutz ratios
for the three values of $\beta$ as follows

\vspace{-0.4cm}

\begin{equation}
\left( { {\bar r} \over a} \right)^2 \chi(R,R) = \sigma {\bar r}^{\ 2} + 2 \gamma 
\left({{\bar r} \over a}\right)^2 {1 \over R^2}+ 4 \left({{\bar r} \over a}\right)^4 {1 \over R^4} \left[ { 1.65 - \sigma {\bar r}^{\ 2} - 2 \gamma \over 4} + d \left({ a \over {\bar r} }\right)^2  \right]
\end{equation}

\vspace{-0.1cm}

\noindent
given in terms of  6 fitting parameters
$\sigma {\bar r}^{\ 2},\ \gamma,\ d,\ {a(\beta=5.96) \over {\bar r}},
\ {a(\beta=6.17) \over {\bar r}},\ {a(\beta=6.42) \over {\bar r}}.$ 
The fit has a reduced chi square of $\chi^2$/ndf=1.05, and the fitted
parameters are 

\vspace{-0.4cm}

\begin{equation}
\label{fitparams}
 \sigma {\bar r}^{\ 2}=1.159(6),\ \gamma=0.250(3),\ d=0.24(2)   
\end{equation}
   
\begin{wrapfigure}{t}{65mm}
\begin{center}
\includegraphics[width=65mm]{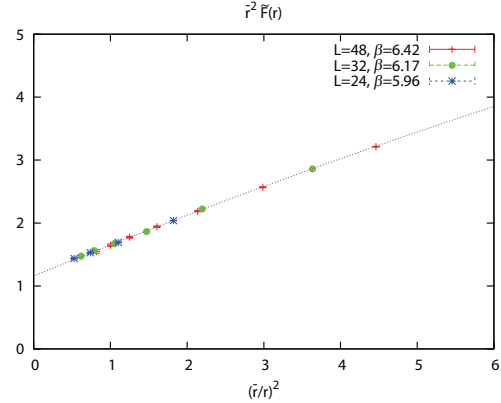}
\end{center}
\vspace{-0.7cm}
\caption{The continuum form of Creutz ratios  ${\bar r}^{\ 2} {\tilde
F}(r)$ plotted as a function of $({\bar r}/r)^2$.}
\label{fig_cratio_cont}
\end{wrapfigure}

\vspace{-0.1cm}   
   
\noindent
plus the values of $a/{\bar r}$  given  in Table 1. 
In Fig. \ref{fig_cratio_phys}  we display the data together with the
fitting function for the three values of $\beta$. Although the three
curves and the data coalesce for large values of $r$, they are clearly
different at smaller $r$. The difference is perfectly accounted for 
by the expected correction to scaling given in terms of the single
parameter $d$. Using the measured value of this parameter we 
display in Fig. \ref{fig_cratio_cont} the result for the continuum function 
$\bar{r}^{\ 2} \tilde{F}(r)$ obtained from our three values of beta.
Scaling is almost perfect. Notice that the resulting function provides
a continuum observable of pure gauge theory which had not been 
measured before. 

The use of the new unit  ${\bar r}$ has allowed us to give a
self-contained determination of the continuum function. However, 
the ratio of scales $a(\beta=6.17)/a(6.42)=1.4249(45)$ is independent 
of the choice of units. This ratio matches perfectly with the
result following from the formula of Ref.~\cite{guagnelli}, which is 
$1.4252$. It also agrees within errors with the value given in
Ref.~\cite{wilsonf}: $1.4310(24)$. The latter  uses the unit $t_0$
mentioned earlier. We have also determined this quantity from our data.
The results are given in Table~\ref{table1}. From it we deduce 
$a(6.17)/a(6.42)=1.4243(21)$. The agreement between our two 
determinations and those of other authors is remarkable. From it we
can determine the ratio of length units $\sqrt{8 t_0}/{\bar{r}}$ which
averages to $0.995(5)$, and using $\sqrt{8
t_0}/{r_0}=0.948(6)$~\cite{wilsonf}, obtain $\bar{r}/r_0$. 

A similar test of scaling can be done by looking at the ratio 
$a(5.96)/a(6.17)$. One expects higher violations of scaling at 
this lower $\beta$ and indeed the results are not so spectacular, 
but well under the percent level.
One gets $1.4123(21)$ from our fit, $1.4063$ from
Ref.~\cite{guagnelli}, $1.4038(24)$ from Ref.~\cite{wilsonf} and 
$1.4038(12)$ from our $t_0$ determination. 

Using the length scale relations  of the previous paragraphs we can
transform our string tension measurement in Eq.~\ref{fitparams}
into other units. We get $\sqrt{8t_0 \sigma}=1.071(6)$ and 
$r_0 \sqrt{\sigma}=1.13(1)$. 
Previous measurements of this quantity  derived from 3-d smeared potential are
almost consistent with the value $r_0 \sqrt{\sigma}=\sqrt{1.65-\pi/12}=1.178$ \cite{3dsm}.
It is not clear how to interpret the 5\% difference between the two values of
$r_0 \sqrt{\sigma}$, but in view of the previous results, it can hardly
be due to errors in the scale determination.
It should be noted, however,  that the two numbers  are obtained from quite different 
geometries of Wilson loops. In fact, for 3-d smearing, we use Wilson loop $W(R,T)$ 
with finite $R$ and infinite  $T$, and the resultant expression of the force is
$r_0^2 F(r) = r_0^2 \sigma  + {\pi \over 12} \left( {r_0 \over r}\right)^2 \simeq r_0^2 \sigma 
+ 0.2612 \left( {r_0 \over r}\right)^2$. 
On the other hand, for 4-d smearing, we use Wilson loop $W(R,T)$ with finite $R \approx T$,
and the resultant expression of the Creutz Ratio is given in eq.~(\ref{eq_cratio_cont}).

In conclusion, we have presented a method based on  4-dimensional Ape smearing
that allows a precise determination of Creutz ratios. We showed that
the results are insensitive to  using Wilson flow instead. The method
is applied to SU(3) pure gauge theory allowing to compute a physical function
of length which is the continuum  equivalent of Creutz ratios, and whose large
size limit is the string tension.

A.G-A acknowledges financial support from the grants FPA2012-31686
and FPA2012-31880, the MINECO Centro de Excelencia Severo Ochoa Program SEV-
2012-0249, the Comunidad Aut\'onoma de Madrid HEPHACOS S2009/ESP-1473, and
the EU PITN-GA-2009-238353 (STRONG net). He participates in the Consolider-
Ingenio 2010 CPAN (CSD2007-00042). M. O. is supported by 
the Japanese MEXT grant No 26400249. 
Calculations have been done on the INSAM clusters at Hiroshima University and 
also on Hitachi SR16000 supercomputer both at High Energy Accelerator Research Organization(KEK) and YITP in Kyoto University.  Work at 
KEK is supported by the Large Scale Simulation Program No.13/14-02.

\end{document}